\documentstyle[epsfig]{aipproc}

\begin{document}

\title{Winds and Shocks in Galaxy Clusters:\\ Shock Acceleration on an
Intergalactic Scale}

\author{T. W. Jones$^*$, Francesco Miniati$^*$,\\ Dongsu Ryu,$^{\dagger}$
and Hyesung Kang$^{\P}$}
\address{$^*$Department of Astronomy, University of Minnesota, Minneapolis, MN 55455, USA\\
$^{\dagger}$Dept. of Astronomy and Space Science, Chungnam National
Univ., Daejeon 305-764, Korea\\
$^{\P}$Department of Earth Science, Pusan National University, Pusan 609-735, Korea}

\maketitle

\begin{abstract}
We review the possible roles of large scale shocks as particle accelerators
in clusters of galaxies. Recent observational and theoretical work
has suggested that high energy charged particles may constitute
a substantial pressure component in clusters. If true that would
alter the expected dynamical evolution of clusters and increase the
dynamical masses consistent with hydrostatic equilibrium. Moderately
strong shocks are probably common in clusters, through the actions of
several agents. The most obvious of these agents include winds from
galaxies undergoing intense episodes of starbursts, active galaxies
and cosmic inflows, such as accretion and cluster mergers. We describe
our own work derived from simulations of large scale structure
formation, in which we have, for the first time, explicitly included
passive components of high energy particles. We find, indeed that
shocks associated with these large scale flows can lead to nonthermal
particle pressures big enough to influence cluster dynamics. These same
simulations allow us also to compute nonthermal emissions from the clusters.
Here we present resulting predictions of $\gamma$-ray fluxes.
\end{abstract}

\section*{Introduction}
%
%
%
%

Clusters of galaxies are very important probes for cosmology, since they are
the largest bound systems in the universe. 
They represent the nonlinear development of large-scale
perturbations in the early Hubble flow.
Matter, energy and entropy can be added from outside through infall or 
cluster mergers, but with the exception of a small fraction of
the energy as photons, neutrinos and ultra high energy cosmic-rays, 
nothing leaves clusters once they form. 
Thus, they provide unique records of the history of the universe.
The statistics of
cluster masses and their dynamical properties, including, for instance,
the relative
proportions of baryonic  and nonbaryonic matter, are commonly used to
test basic cosmological models \cite{bah99}. While galaxies
are the most obvious constituents of clusters in visible light,
most of the matter in clusters is nonbaryonic, and even the baryonic
matter is primarily contained within the diffuse intracluster medium (ICM),
rather than the galaxies.
The temperature and density distribution of the ICM gas directly reflects
the dynamical state of the clusters, as well as holding a history of
important dissipative processes encountered by the gas \cite{loew00}.
Elemental abundances
in the ICM are seen as key indicators of the star formation histories 
and of galaxy evolution, more generally \cite{mush99,volk96}. 
The dynamical states of clusters have also received much attention.
While cluster ICMs sometimes appear relaxed, 
often the situation is quite different, with clear
indications of high speed flows \cite{don98,dup00,kiku00,knopp96,mark99} 
demonstrating that cluster environments can be violent.

The possible importance of nonthermal components in the ICM has
recently raised great interest. There is growing evidence that
magnetic fields \cite{clarke00,kron00}, and high energy charged particles
\cite{blasi99,kaas99,lieu99} may constitute significant dynamical components of
the ICM in at least some clusters.
If these components are generally
strong, they would impact on a wide range of issues, beginning with
estimates of cluster masses derived from assumptions of ICM hydrostatic
equilibrium. In addition, since high energy charged particles and magnetic
fields do not readily radiate away their pressures, they could tend to
inhibit cooling flows. Radio, EUV and hard X-ray emissions resulting from 
energetic electrons have already stimulated much discussion about cluster physics as
well as the evolution of the clusters and their galaxies \cite{ber97,enss97,kron99,roet99,sar98,volk96}.
While high energy $\gamma$-rays have not yet been detected from clusters,
recent estimates of {$\gamma$-ray} luminosities from high energy
particle interactions in the nearest rich
clusters, such as Coma, are within the range of what may be detected in the
next generation of $\gamma$-ray observatories \cite{blasi99,cola98}.

Our focus here is the possible relationships between shock waves generated
by high speed flows in clusters and the origins of the high energy charged
particles. For convenience, we call all such particles cosmic-rays or ``CRs''.
A few basic considerations of the CRs are necessary to set the stage. So far
direct observations reveal only the presence of CR electrons, although one generally
assumes that protons are present with at least comparable numbers and
greater energy content \cite{lieu99}.
Relativistic protons below energies where 
photo-pion production through interaction with the cosmic
microwave background (CMB) becomes important,
($\sim 10^{9.5}$ GeV) do not lose significant fractions of their energy
in a cluster over a Hubble time \cite{ber97}. Also, up to somewhat lower
energies they should scatter off magnetic irregularities sufficiently to remain
trapped in the cluster for a similar duration \cite{volk96}.
Relativistic electrons, on the other hand can only survive Coulomb losses
and radiative losses for more than $\sim 1$ Gyr in a relatively
narrow energy band centered roughly around 50 MeV \cite{sar99}.
Thus, observed relativistic electrons must be freshly accelerated,
reaccelerated or be introduced as 
decay products of other interactions \cite{brun00,dol00,eil99}.

The Coma cluster provides a convenient scale for the energetics involved
in producing and containing the CRs. The dynamical mass estimates of
Coma are a bit over $10^{15}~M_{\sun}$ \cite{gell99}, leading to an
ICM mass $\sim 10^{14}~M_{\sun}$. With a mean ICM temperature of 8 keV \cite{wat99}
the total thermal energy in the ICM $\sim 4\times 10^{63}$ erg. Observed
excess EUV emission from Coma, interpreted as inverse Compton scattered CMB photons,
leads to an estimated $\gtrsim 10^{61}$ erg in total CR electron energy and,
through conventional arguments, $\gtrsim 10^{63}$ erg in protons 
\cite{lieu99,sar98}. If we accept that figure for argument and assume the
CR protons were supplied continuously over a Hubble time the mean CR
energy input rate would be $\sim 3\times 10^{45}$ erg/sec. By comparison
the energy input to CRs in our own galaxy is only about 
$10^{41}$ erg/sec \cite{drury89,field00}.
This estimate of CR energy in Coma makes it clearly very important
dynamically, and the implied rate of energy input severely restricts
the range of possible accelerators.
\begin{figure}[b!] 
\centerline{\epsfig{file=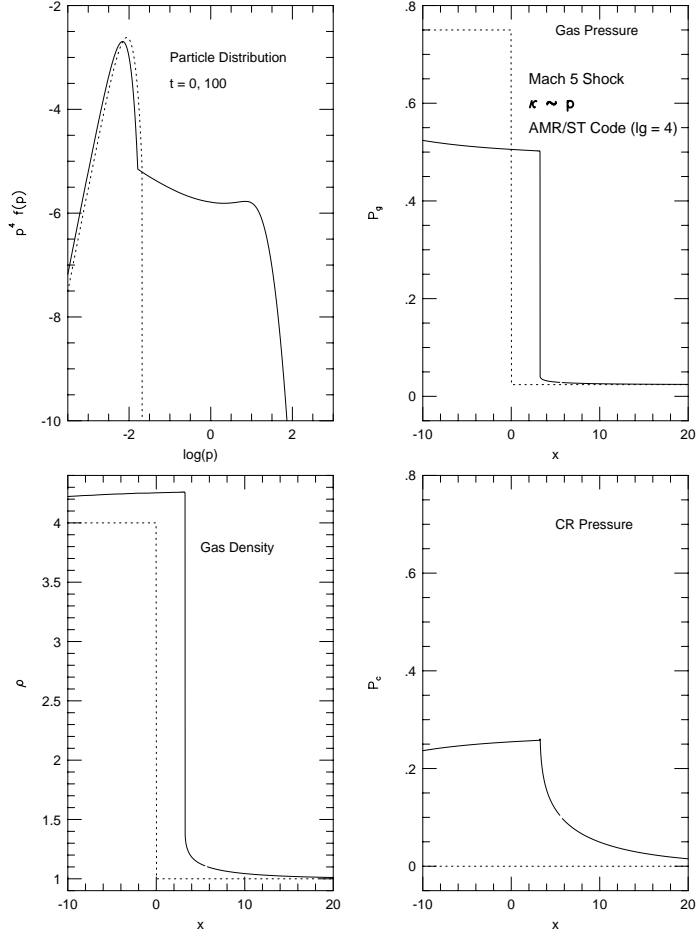,height=6.0in,width=4.6in}}
\vspace{10pt}
\caption{Properties of a simulated Mach 5 shock including backreaction
from nonthermal particles accelerated by DSA after being injected
through ``thermal leakage''. The momentum distribution of the particles
just downstream of the shock is also shown. The dotted lines show
initial conditions, while the solid/dashed lines show properties
after CR protons have been accelerated to energies $\sim 100$ GeV
and the shock structure is almost steady.
Bohm-like spatial diffusion, $\kappa \propto p$, is assumed.}
\label{fig1}
\end{figure}

\section*{Large-Scale Shocks as Accelerators}

There are likely to be many particle accelerators in clusters, but only a 
few seem really capable of accounting for the CR energy content
implied above for clusters like Coma. Thus, for example, while second order Fermi
acceleration from MHD turbulence in the ICM might conceivably
play a role in electron acceleration \cite{brun00,eil99}, that would
not seem to be a very fruitful approach to accounting for 
proton CRs, if their energy content exceeds $\sim 10^{63}$ erg in
rich clusters. In fact, most current
models for proton acceleration depend on first order Fermi acceleration
resulting from CR diffusion near shocks. This so-called ``diffusive
shock acceleration'', or DSA, can be efficient enough to put into CRs
several tens of percent of the total energy flux through a shock \cite{ber00}.
Following that lead, the initial task becomes one of identifying shocks that
dissipate enough energy to explain the energetics suggested for Coma
and by implication other rich clusters. It turns out that large scale shocks 
are probably fairly common in ICMs, so it may indeed be reasonable
to expect a substantial energy in CR protons there. Electrons
are more difficult to accelerate and maintain than protons. First,
they radiate energy much more rapidly \cite{sar99}. In addition,
they are harder to get started in most accelerators,
because of their small gyroradii at suprathermal energies.
Still, from galactic CRs and other environments, such as radio galaxies,
we have abundant evidence that electrons can be accelerated. Also,
inelastic proton-proton interactions with the ICM can generate a 
significant secondary electron-positron component through
pion decay \cite{den80,dol00}.

Several candidate classes of large shock structures come to mind immediately
in clusters.
For example, the facts that ICM gas is generally enriched with 
nucleosynthetic products to $10-30$\% of the solar value \cite{mush99}
and that starburst galaxies like M82 produce
strong winds, have been cited to support the idea that termination
shocks in galactic winds may have been common enough during early
``bright phases'' of galaxy evolution to 
accelerate a substantial high energy CR population \cite{jok87,volk96}.
Starburst-like energy deposition rates are probably necessary to
generate winds inside rich clusters, in order
to overcome the substantial thermal pressure of the ICM \cite{volk96}.
If one takes an ICM Fe mass $\sim {\rm few}\times 10^{11} M_{\sun}$ 
\cite{bohr96} and assumes it resulted from
supernovae, then $\sim 10^{12}$ events, corresponding to about $10^{63}$ erg
of kinetic energy would have been released into a rich cluster like Coma
\cite{volk99}.
That is close to the CR energy content required, but does not yet
account for a realistic conversion efficiency. Even if we assume $\sim$100\%
of the supernova energy ends up as kinetic energy in the winds and that we can carry over
from galactic supernova a CR acceleration efficiency 
$\sim 10~-~20$\% \cite{heck00}, this model seems to struggle to account for the very large
energy contents if the recent EUV observations represent
inverse Compton emission and the implied CRs fill the entire cluster \cite{lieu99}.
On the other hand, these winds may still be effective accelerators
or reaccelerators at a somewhat lower level, as well as contributors
to the ICM magnetic field \cite{kron99}.

CRs escaping from radio galaxies have long
been seen as potential sources of relativistic particle populations
in clusters and sometimes beyond \cite{enss97,jaff77}. 
These CRs are usually seen to be 
accelerated by the terminal shocks of the jets powering the
radio lobes, but it also seems plausible that the bow shocks of at least
the most powerful radio jets in cluster environments might be effective 
accelerators.
Energetically, radio galaxies come close to required input levels if we can
assume on average that a rich cluster contains at least one at all times
\cite{ber97}. 
The most luminous radio galaxies are estimated to be driven by jet
kinetic powers $\gtrsim 10^{45}$ erg/sec \cite{beg84}, which is comparable
to our earlier estimated average rate of CR production for Coma. 
Individual radio galaxy lifetimes are generally
estimated at $\sim 10^{8}$ years \cite{beg84}, so such a cluster
would need as many as $\sim 10^2$ luminous galaxies capable of
generating a powerful radio jet or a fairly high duty cycle in a
smaller number of galaxies.
On
the other hand, this localized source model 
may have some difficulty explaining the diffuse radio halos seen in
some luminous clusters \cite{liang99}, including Coma, 
because of large energy loss rates for high energy
electrons and because low energy electrons probably cannot diffuse very 
far from their source \cite{jaff77}.
Reacceleration can, of course, reduce or eliminate this issue \cite{brun00}.
Similarly, if the energy were deposited in high energy protons the difficulty
is reduced. Then the nonthermal emissions observed so far would presumably
be explained through secondary electrons.
There are also suggestions that the relatively light plasma in radio lobes
may remain confined in buoyant ``ghost'' bubbles, so that they can be reenergized
and illuminated at later times by passage of other large scale 
shocks \cite{enss99}.

Another scenario gaining some strong
recent support is CR acceleration at very large and long-lived
shocks resulting from cosmic structure formation. The likely existence
of strong accretion shocks several megaparsecs from cluster cores
has been recognized for a long time \cite{bert85,rk97b}, and they have been
suggested as possible sources of ultra-high energy CRs \cite{kang96,kang97}
as well as seeds for the ICM magnetic field \cite{kcor94}. Shocks resulting
from discrete cluster merger events have also been recognized through
X-ray structure in clusters \cite{don98,mark99}, and cited as particle
accelerators to account for diffuse cluster radio halos \cite{har80,tribb93}
or so-called ``relic'' radio sources \cite{roet99,rott97}.
The overall energetics of these shocks is generally acceptable for
the production of the CRs needed. Typical flow speeds in and around
clusters will be $v_f \sim (2G~M_{cl}/R_{cl})^{1/2} \sim 2\times 10^3$ km/sec,
leading to an available power $\sim \rho_{cl} v^3_f R^2_{cl} \sim 10^{46}$ erg/sec,
using $M_{cl} \sim 10^{15} M_{\sun}$ and $R_{cl} \sim 2$ Mpc.
Accretion shocks far from cluster cores can be of very high Mach number
and are responsible for initial heating of the ICM.
They appear, however, to be less important as potential sources of
CRs inside clusters than weaker, ``internal'' shocks associated
with mergers and other flows penetrating deeper into the clusters. 
The reason is that the latter shocks repeatedly process the ICM
material, whereas the accretion shocks do it only once. 
\begin{figure}[b!] 
\centerline{\epsfig{file=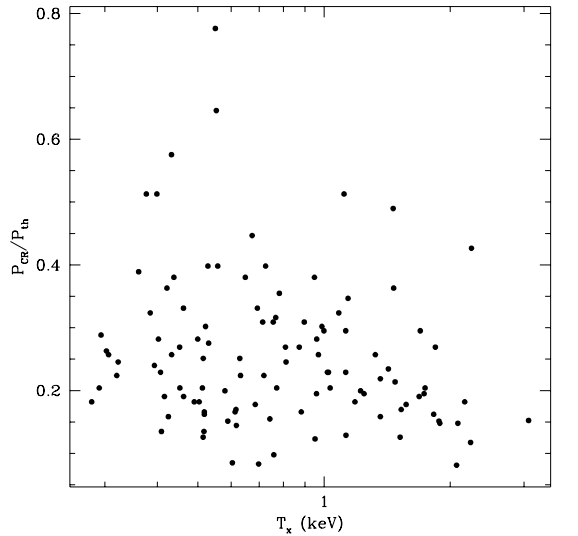,height=3.0in,width=3.0in}}
\vspace{10pt}
\caption{Mean $P_{CR}/P_{th}$ measured in simulated clusters inside
$r = 0.5$ Mpc/h at $z=0$.}
\label{fig2}
\end{figure}

Miniati
{\it et al} \cite{min00} explored the statistics  of shock  heating
in both SCDM and $\Lambda$CDM cosmologies. They found that
far more mass gets processed through the internal shocks than the
accretion shocks. Their analysis showed that the most common shock
encounters by the ICM material involve shock Mach numbers less than
10, with the peak around $M \sim 5$. That is significant, since such
shocks are strong enough to transfer as much as 20~-~30\% of the postshock
pressure to CRs. On the other hand such shocks are only
modestly modified in their structures through backreaction of CRs diffusing upstream.
Thus, so-called ``test particle'' estimates of the CR distributions
are a reasonable first guess to what we may expect.
Figure 1 illustrates the evolution of a CR modified Mach 5 shock
as computed from a fully nonlinear
DSA simulation carried out using methods detailed elsewhere \cite{kang00}.

\section*{Simulating CR Acceleration in Cosmic Structure Formation}

So far, only a few clusters have shown direct evidence of large shocks. 
However, recent cosmic structure formation simulations have demonstrated
that large scale shocks in clusters are probably much more common and much more
complex than these simple perceptions suggest \cite{min00}.
Since clusters tend to form at the intersections of cosmic filaments, 
they accrete matter in very unsteady and nonisotropic patterns. In addition to
discrete cluster mergers, general, larger scale flows 
and shocks associated with them propagate down the filaments and through
the clusters. When cluster mergers take place the accretion shocks
associated with the individual clusters add to the shocks
that form in direct response to the merger.
The net result of all of this is a rich web of shocks, which often
penetrates into the inner regions of the clusters.
\begin{figure}[b!] 
\centerline{\epsfig{file=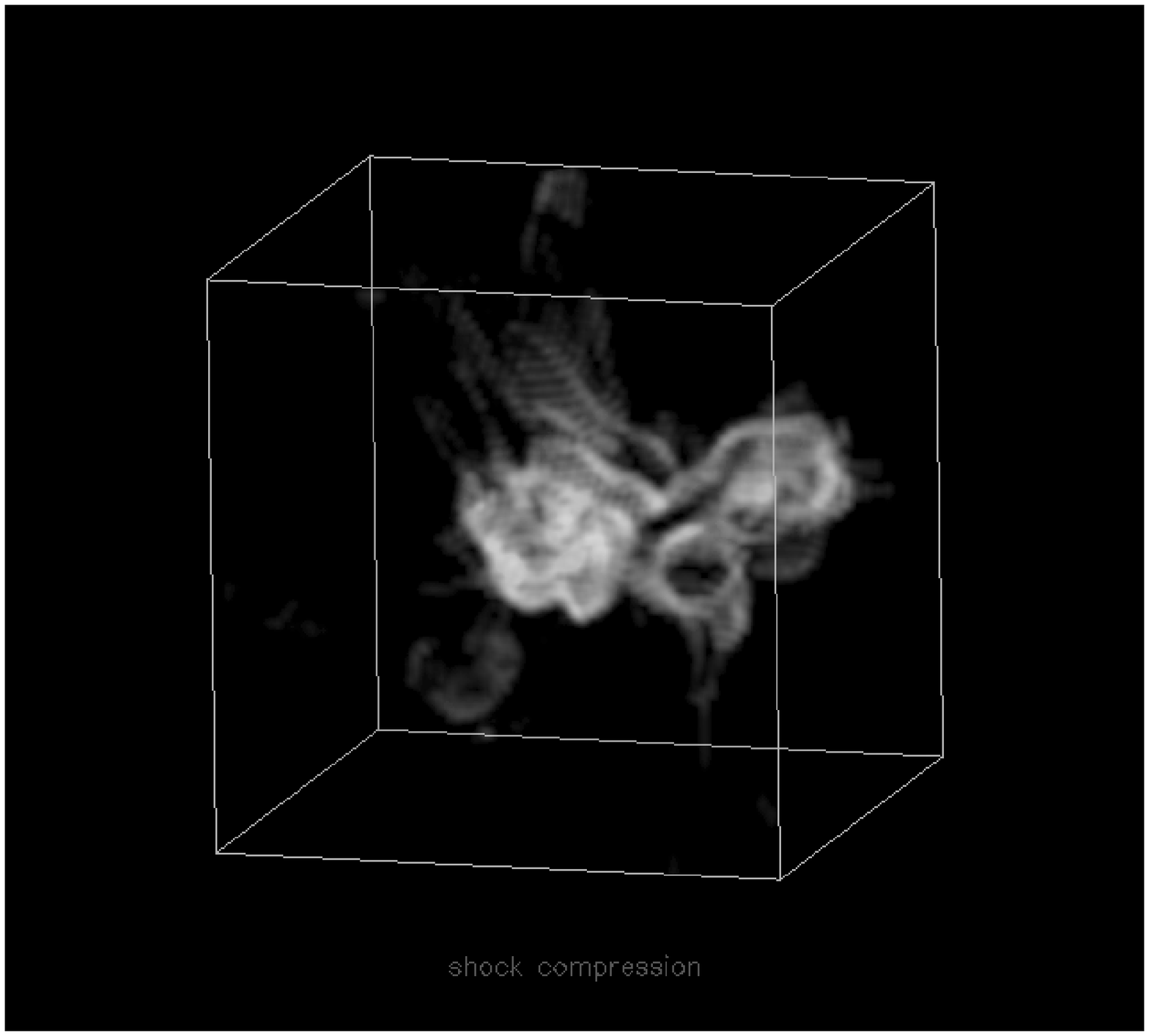,height=2.4in,width=2.3in}
\epsfig{file=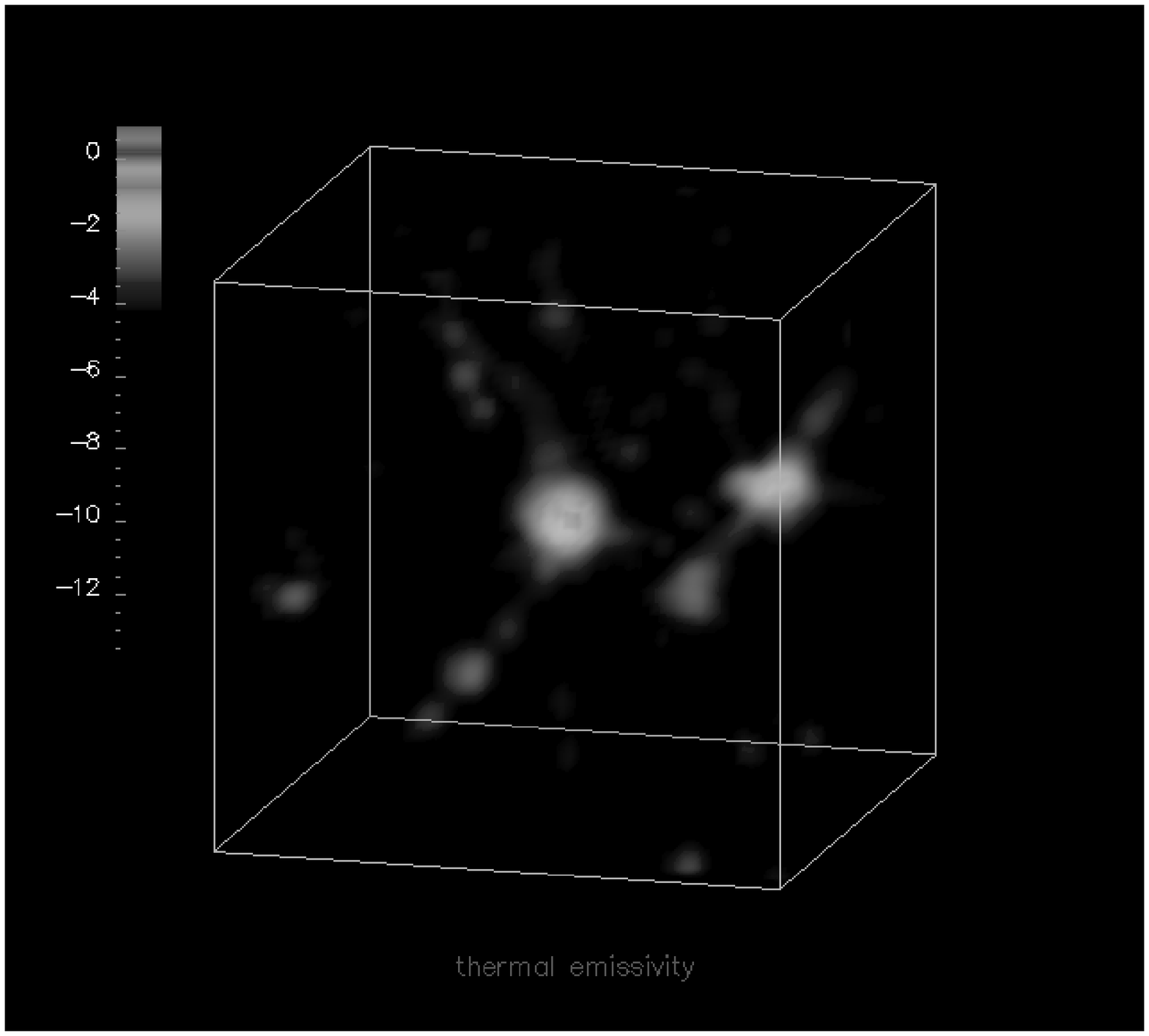,height=2.4in,width=2.3in}}
\vspace{10pt}
\caption{Volume renderings of clusters formed in a SCDM 
simulation. 
Left: Shock surfaces. Right: Thermal
bremsstrahlung emissivity. A $14$ Mpc/h$^3$ volume is shown at z = 0
from a $50$ Mpc/h$^3$ simulation.}
\label{fig3}
\end{figure}

The recent detections of shocks from merger events has been noted
above. But the existence of accretion shocks is not necessarily easy to
detect directly. On the other hand, infalling clouds of unshocked, warm 
gas outside clusters
may be identified through the absorption lines of quasars
located inside clusters.
This warm low density gas of $10^4-10^5$K is photoionized by the 
diffuse radiation from the hot postshock gas and the diffuse cosmic 
background radiation. In some studies \cite{fol86,wey79},
the C IV absorption systems of quasar emission lines with 
$ |z_{abs}-z_{QSO}| < 3000~{\rm km~s^{-1}}$ are interpreted as
clouds associated with rich clusters where the quasars reside. 
The characteristics of these systems of C IV absorbers are different from
those of the typical intervening C IV absorbers.
It has been noted that the velocity difference is unexpectedly large 
compared to the typical velocity dispersion of galaxies ($400-1200{\rm km~s^{-1}}$)
in rich clusters \cite{fol86}.  
An accretion velocity, however, is a bit larger than the galaxy velocity
dispersions, since it is given by $v_{acc}= 1.31\times 10^3 {\rm km~s^{-1}}
(T_{cl}/6.06 {\rm keV})^{1/2}$, for example, in an Einstein-de Sitter universe
\cite{bert85,rk97}.
Thus it is possible that these absorption systems are in fact the 
infalling clumps of gas upstream of the accretion shock.

Large shocks may be very efficient sources of CRs in clusters.
We now outline a computational
study of this phenomenon as revealed in a simulation by Miniati \cite{minth}.
Briefly, this calculation used an Eulerian ``TVD'' hydro~+~N-body cosmology code 
\cite{ryu93} with passive treatments of magnetic fields \cite{kcor94} and CRs added.
The CR protons and electron populations were accelerated according to DSA
test-particle theory, then evolved to include adiabatic, radiative and
Coulomb energy losses up to $10^6$ GeV, using methods similar
to those described previously \cite{jones99}. The upper energy limit was chosen
so spatial diffusion could be neglected in order to
reduce computational costs, and since most observable emissions can be
studied in this regime. 

Secondary electron production due to pion
decays, was also included. CRs were injected at shocks according
to the ``thermal leakage''  model described by Kang \& Jones \cite{kang95}.
Roughly speaking, a small fraction of the downstream thermal particles, whose
velocities are large enough for them to escape back across the shock, are assumed to be
``injected'' into the nonthermal particle population, whence they are
subject to DSA.
A 50 Mpc/h$^3$ periodic section of
a SCDM universe was simulated in this case on a $256^3$ grid, with
h = 0.5, $\Omega_M = 1$, $\Omega_B = 0.13$ and $\sigma_8 = 0.6$.
\begin{figure}[b!] 
\centerline{\epsfig{file=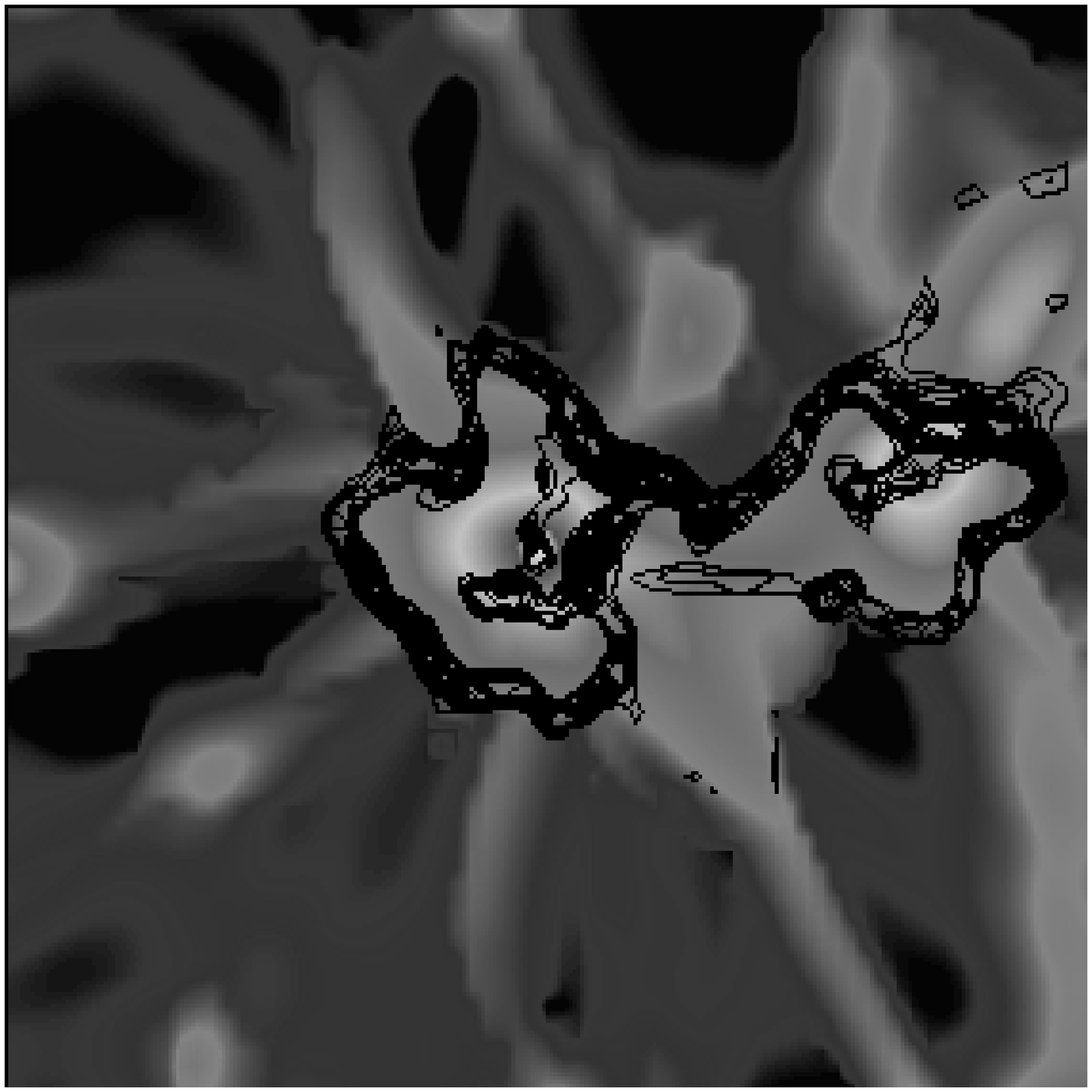,height=2.4in,width=2.3in}
\epsfig{file=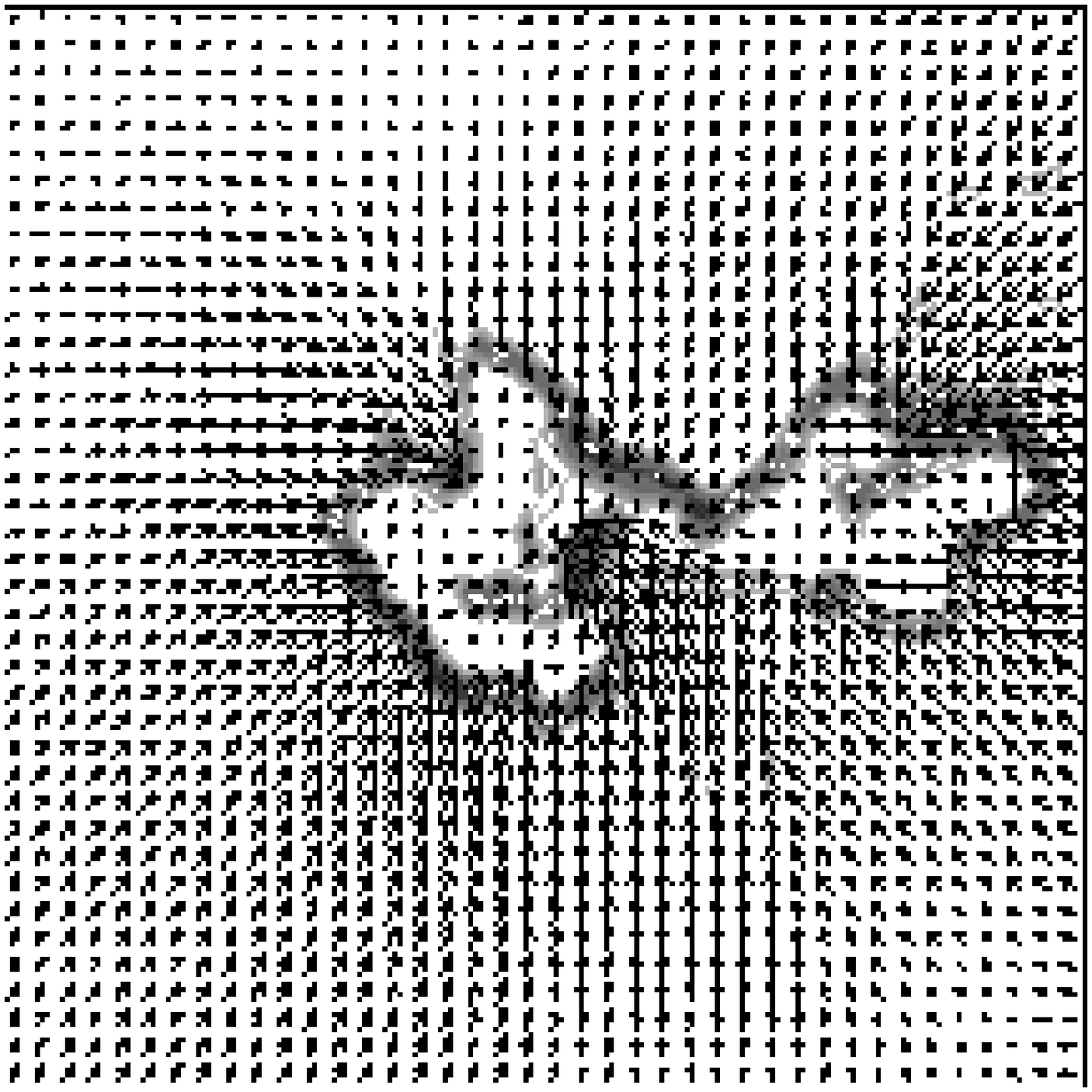,height=2.4in,width=2.3in}}
\vspace{10pt}
\caption{Vertical slices through the clusters shown in Figure 1. Left:
Thermal bremsstrahlung emissivity with shock surfaces superposed.
Right: Shock surfaces superposed on the projected flow velocity 
field.}
\label{fig4}
\end{figure}
\begin{figure}[b!] 
\centerline{\epsfig{file=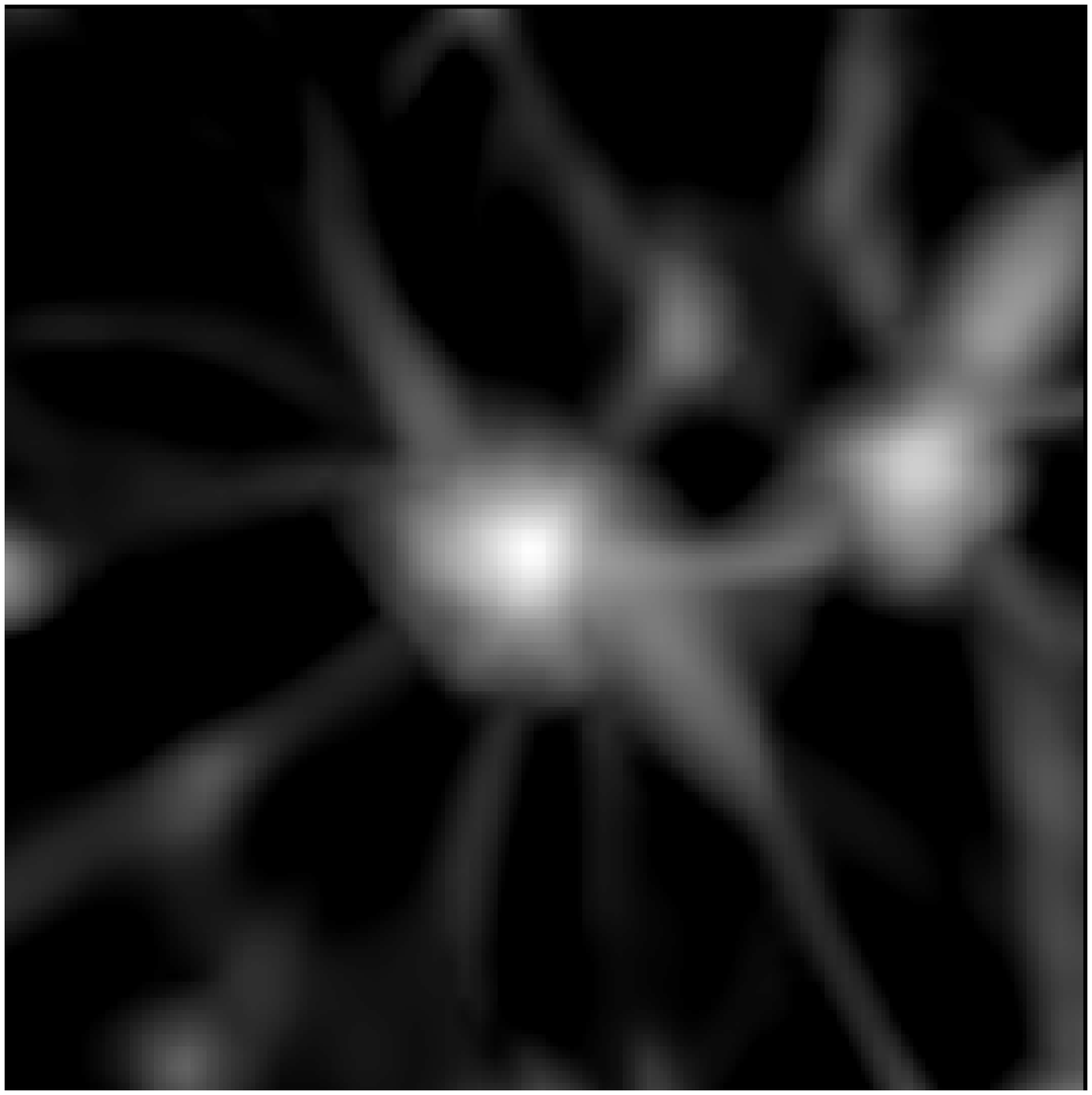,height=2.4in,width=2.3in}
\epsfig{file=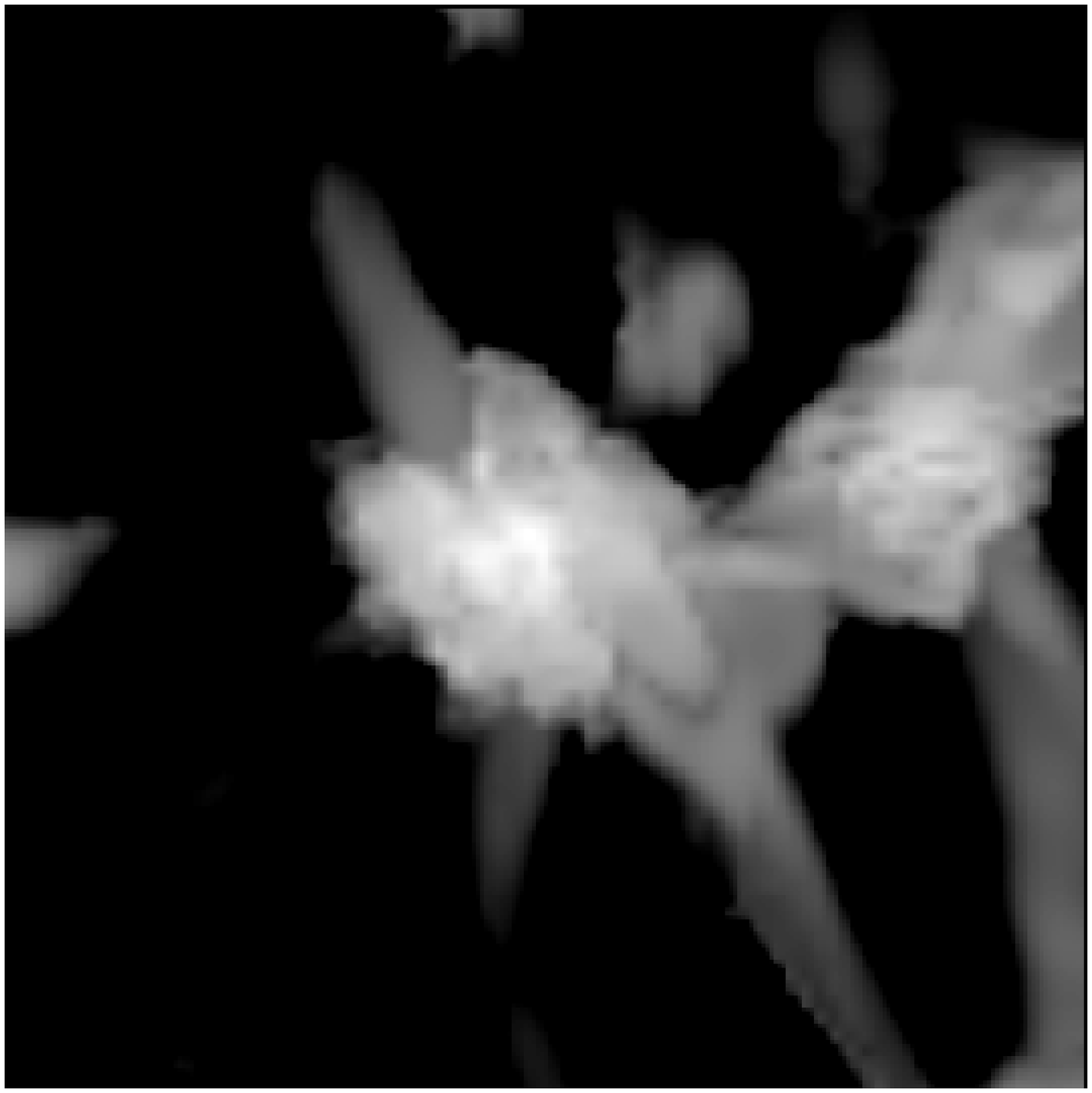,height=2.4in,width=2.3in}}
\vspace{10pt}
\caption{Vertical slices as in Figure 3. Shown are Left: Log thermal proton
density. Right: Log CR proton density.}
\label{fig5}
\end{figure}

Figure 2 shows the mean ratio of CR 
pressure to thermal pressure in the central regions of clusters in
this simulation. These results are indicative only, since the
CRs were not coupled back dynamically in cluster evolution. The mean
ratio, $P_{CR}/P_{th} \sim 0.25$, however, consistent with the 
results of nonlinear Mach 5 CR shock simulations 
mentioned earlier. There is considerable scatter, since
each cluster has a unique shock history.
These nonthermal pressures are high enough that
they would produce non-negligible influences on the internal
cluster dynamics. Clearly, future simulations should consider those
effects self-consistently.

The shock structures associated with the clusters are very complex.
Figure 3 isolates from the simulation a 14 Mpc/h$^3$ cube at z = 0, 
centered on a pair of interacting clusters. It shows clearly
the web of shock surfaces in and around the
clusters. The manner in which these shocks have formed and the considerable degree
to which they penetrate into the cluster cores is shown in Figure 4.
Here we have taken a vertical slice through the clusters in Figure 3, 
and show the intersections of the shock surfaces (contours of $\nabla\cdot v$)
with the slice plane. These are superposed onto the thermal bremsstrahlung
X-ray emissivity and projections of the vector velocity field. Associated
cosmic filamentary features are evident, as well as the fact that
strong flows are focussed along the filaments and directed at the 
clusters. Where those flows enter the clusters shocks sometimes  penetrate
deeply into the cores.
\begin{figure}[b!] 
\centerline{ \epsfig{file=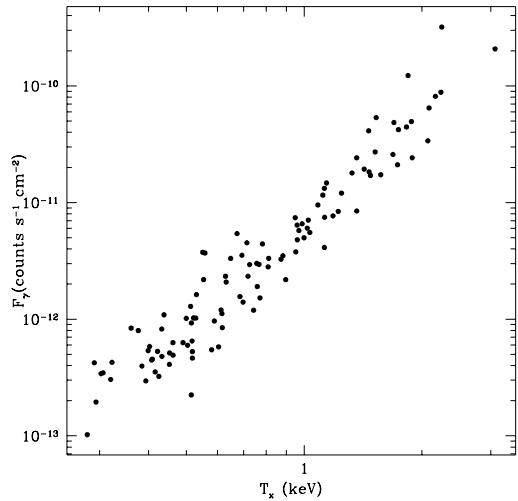,height=2.8in,width=2.8in}}
\vspace{10pt}
\caption{High energy $\gamma$-ray flux ($>$ 100 MeV) from simulated 
clusters at $z=0$ assuming a distance 70Mpc/h, with h = 0.5.}
\label{fig6}
\end{figure}

Figure 5 provides a comparison of gas and CR distributions 
in the same plane. On the left we show the spatial distribution of
thermal protons, and on the right the distribution of CR protons.
While there is a crude correspondence between the
CRs and the gas, they clearly are not identical.
The most obvious distinction is the existence of sharp edges in
the CR distribution,
reflecting their injection at shocks, while the thermal density
increases smoothly towards cluster centers. The rate of $\gamma$-ray
production from pion decay depends on the collision rate between 
thermal and CR protons, so scales as the sum of these two (logarithmic)
images. Thus, the $\gamma$ ray emission may have extensions
beyond the thermal X-rays, and may show fine structure not evident in the
thermal emission, as well. As mentioned earlier, most of the shock
processing of the gas in clusters involves shocks with Mach numbers 
$\sim 5$ or so. Thus, the computed CR energy distributions typically 
resemble $N(E) \propto E^{-2.1}$,
and the predicted $\gamma$-ray spectrum has a similar form.

In Figure 6 we show the simulated $\gamma$-ray flux, $F_{\gamma}(> 100 {\bf MeV})$, emitted 
within 1.3 Mpc/h of cluster centers as a function of the cluster
temperature, $T_x$, assuming 
each is at the distance of the Coma cluster, 70 Mpc/h with h = 0.5. 
The least squares fit to this distribution gives $F_{\gamma} \propto T_x^{2.95}$.
A scaling close to $F_{\gamma} \propto T_x^{3}$ makes sense in this
model from the
fact that the kinetic power in accretion shocks of virialized
clusters scales as $T_x^2$, providing a scaling for $P_{CR}$ or $n_{CR}$,
while the mean cluster baryon densities
scale approximately as $T_x$ in the simulation. 
The $F_{\gamma}$:$T_x$
relation found in this calculation is considerably steeper
than in some other models, such as that by Colafrancesco \& Blasi \cite{cola98},
which assumes all the CR protons diffuse from a central injection point
and predicts closer to $F_{\gamma} \propto T_x^{0.5}$.
Our simulation utilized a box too small to produce clusters as
massive as Coma, but an extrapolation of the $\gamma$-ray
relation in Figure 6 to $T_x = 8.3$ keV leads to a prediction of
about $3 \times 10^{-9}$ counts/sec/cm$^2$, roughly the same as
the value found by Colafrancesco and Blasi for the
same cluster mass, despite the other differences in the models. 
The current EGRET limit
for Coma is $4\times 10^{-8}$ \cite{sre96}, so an order of magnitude
higher. Still, the coming generation of $\gamma$-ray telescopes
should be able to test such models as this. Indeed, those telescopes
promise a genuine opportunity to probe the physics of galaxy clusters
in telling ways not possible before.

The work of TWJ and FM has been supported in part by NASA grant NAG5-5055,
by NSF grants AST96-16964 and AST00-71167, and by the University of Minnesota
Supercomputing Institute.
FM also wishes to acknowledge a Doctoral Dissertation Fellowship
from the University of Minnesota.
DR and HK were supported in part by grant 1999-2-113-001-5 from the
interdisciplinary Research Program of the KOSEF.

\end{document}